\documentstyle[sprocl,psfig]{article}
\def\Journal#1#2#3#4{{#1} {\bf #2}, #3 (#4)}
\def\NPB{{\em Nucl. Phys.} B}

\def\PRD{{\em Phys. Rev.} D}

\def\mco{\multicolumn}

\def\ra{\rightarrow}

\def\ko{K^0}

\def\be{\begin{equation}}
\def\ee{\end{equation}}
\def\bea{\begin{eqnarray}}
\def\eea{\end{eqnarray}}
\begin{document}
\title{RENORMALIZATION GROUP ANALYSIS OF THE HEAVY FIELD 
DECOUPLING IN A SCALAR THEORY }
\author{ A. BONANNO, J. POLONYI }
\address{Laboratory of Theoretical Physics and Department of Physics,
\\Louis Pasteur University, 67087 Strasbourg Cedex, France}  
\author{ D. ZAPPAL\`A}
\address{Dipartimento di Fisica, Universit\`a di Catania, and INFN Sezione di 
Catania,\\
Corso Italia 57, 95129 Catania, Italia}
%%%%%%%%%%%%%%%%%%%%%%%%%%%%%%%%%%%%%%%%%%%%%%%%%%%%%%%%%%%%%%
% You may repeat \author \address as often as necessary      %
%%%%%%%%%%%%%%%%%%%%%%%%%%%%%%%%%%%%%%%%%%%%%%%%%%%%%%%%%%%%%%
\maketitle\abstracts{
The problem of  two coupled scalar fields, one with mass much lighter than 
the other is analysed by means of Wilson's renormalization group approach.
Coupled equations for the potential and the wave function renormalization
are obtained by means of the gradient expansion.
The Appelquist-Carazzone theorem is checked at energies comparable with the 
light mass; the physics of the light field is also studied at energies
comparable and greater than the heavy mass.}

%\section{Guidelines}
%\subsection{Producing the Hard Copy}\label{subsec:prod}

It is well known that the effects on the physics of a field, 
due to a much heavier field coupled to the former, are not detectable at
energies comparable to the lighter mass. More precisely the 
Appelquist-Carazzone (AC) theorem~\cite{ac} states that  for a Green's function
with only light external legs, the effects of the heavy loops
are either absorbable in a redefinition of the bare couplings
or suppressed by powers of $k/M$ where $k$ is the energy scale
characteristic of the Green's function 
(presumably comparable to the light mass), 
and $M$ is the heavy mass. However the AC theorem does not allow
to make any clear prediction when $k$ becomes close to $M$ and, in this 
region one should expect some non-perturbative effect due to the onset of new
physics.
\par 
In the following we shall make use of the Wilson's renormalization
group (RG) approach to discuss the physics of the light field from the 
infrared region up to and beyond the mass of the heavy field.
Incidentally, the RG 
technique has been already employed to proof the AC theorem~\cite{girar}.
The RG establishes the flow equations of the various coupling constants of
the theory for any change in the observational energy scale; moreover 
the improved RG equations, originally derived by F.J. Wegner 
and A. Houghton~\cite{weg},
where the mixing of all couplings (relevant and irrelevant) generated 
by the blocking procedure is properly taken into account, should
allow to handle the non-perturbative features arising 
when the heavy mass threshold is crossed.
\par
We shall discuss the simple case of two coupled scalar fields and since 
we are interested in the modifications of the parameters 
governing the light field, due to the heavy  loops, we shall consider 
the functional integration of the heavy field only. The action at a given 
energy scale $k$ is
\be
S_k(\phi,\psi)=\int d^4 x~\left ({1\over 2} \partial_\mu \phi 
\partial^\mu \phi+
{1\over 2} W(\phi,\psi) \partial_\mu \psi \partial^\mu \psi+ 
U(\phi,\psi) \right ),
\label{eq:acteff}
\ee
with polynomial $W$ and $U$ 
\be
U(\phi,\psi)=\sum_{m,n}{{G_{2m,2n} \psi^{2m}\phi^{2n}}\over {(2n)!(2m)!}};
~~~~~~~~~~~~~~~~~~~~
W(\phi,\psi)=\sum_{m,n}{{H_{2m,2n}\psi^{2m}\phi^{2n}}
\over {(2n)!(2m)!}}.
\label{eq:svil}
\ee
Since we want to focus on the light field, which we choose to be $\psi$,
we have simply set to 1 the wave function renormalization of $\phi$.
In the following we shall analyse the symmetric phase of the theory 
with vanishing vacuum energy $G_{0,0}=0$.
\par 
We do not discuss here the procedure employed~\cite{jan} 
to deduce the RG coupled equations for the
couplings in Eq.~\ref{eq:svil}, because it is thoroughly explained in the 
quoted reference. Since it is 
impossible to handle an infinite set of equations and a truncation in the 
sums in Eq.~\ref{eq:svil} is required, we keep in the action
only terms that do not exceed the sixth power in the fields and their 
derivatives. Moreover we choose the initial condition for the RG equations
at a fixed ultraviolet scale $\Lambda$ 
where we set $H_{0,0}=1$, $G_{0,4}=
G_{2,2}=G_{4,0}=0.1$ and
$G_{0,6}=G_{2,4}=G_{4,2}=G_{6,0}=
H_{2,0}=H_{0,2}=0$,
and the flow of the various couplings is determined as a function of
$t=ln \left (k/ \Lambda\right )$, for negative $t$.
\begin{figure}
\psfig{figure=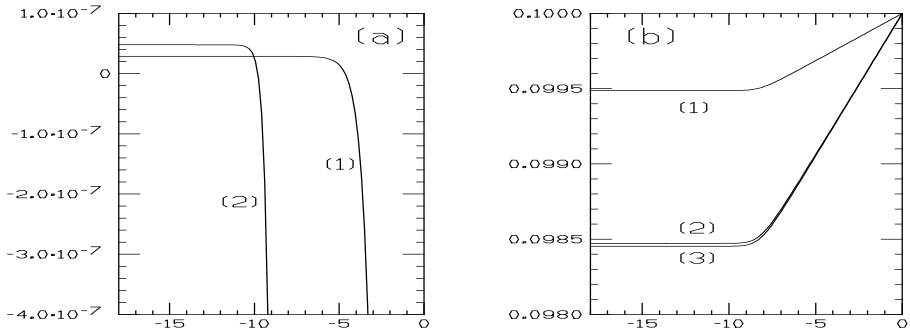,height=4.3cm,width=12.cm,angle=90}
\caption{
(a): $G_{0,2}(t)/\Lambda^2$ (curve (1)) and $10^6\cdot
G_{2,0}(t)/\Lambda^2$ (curve (2)) vs $t=log\left ({{k}/{\Lambda}}\right )$.
\break
(b): $G_{2,2}(t)$ (1), $G_{0,4}(t)$ (2),
$G_{4,0}(t)$ (3) vs $t$.
\label{fig:funo}}
\end{figure}
\par
\begin{figure}
\psfig{figure=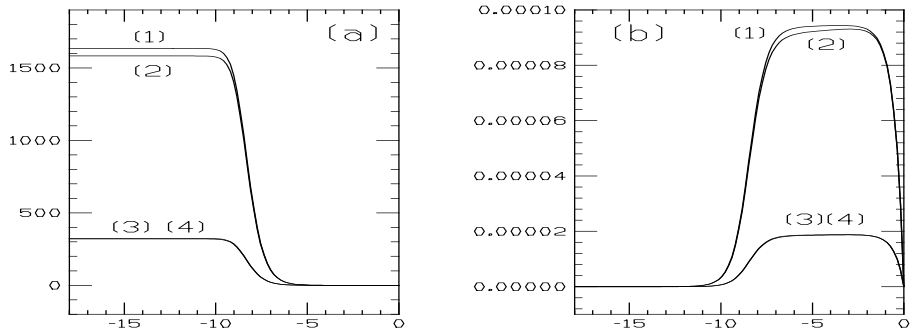,height=4.3cm,width=12.cm,angle=90}
\caption{
(a): $G_{6,0}(t)\cdot \Lambda^2$ (1), $G_{0,6}(t)\cdot \Lambda^2$ (2), 
$G_{4,2}(t)\cdot \Lambda^2$ (3) and 
$G_{2,4}(t)\cdot \Lambda^2$ (4) vs $t$.\break
(b): $G_{6,0}(t)\cdot k^2$ (1), $G_{0,6}(t)\cdot k^2$ (2), 
$G_{4,2}(t)\cdot k^2$ (3) and 
$G_{2,4}(t)\cdot k^2$ (4) vs $t$.
\label{fig:fdue}}
\end{figure}

In Fig.~\ref{fig:funo}(a) $G_{0,2}(t)/\Lambda^2$ (curve (1) ) and 
$10^6 \cdot G_{2,0}(t)/\Lambda^2$ (curve (2)) are plotted. Clearly the heavy 
and the light masses become stable going toward the IR region and their value 
at $\Lambda$ has been chosen in such a way that the stable IR values are, 
$M\equiv\sqrt {G_{0,2}(t=-18)}\sim 10^{-4}\cdot \Lambda$
and $m\equiv\sqrt{G_{2,0}(t=-18)}\sim 2\cdot 10^{-7}\cdot\Lambda$.
So, in principle, there are three scales: 
$\Lambda$, ($t=0$), the heavy mass $M$, 
($t\sim -9.2$), the light mass $m$, ($t\sim -16.1$).
In Fig.~\ref{fig:funo}(b) the three renormalizable dimensionless 
couplings are shown; the neat change around $t=-9.2$, that is $k \sim M$, 
is evident and the curves become flat below this value.
The other four non-renormalizable couplings included in $U$ are 
plotted in Fig.~\ref{fig:fdue}(a), in units of $\Lambda$.
Again everything is flat below $M$, and the values of the couplings 
in the IR region coincide with their perturbative Feynman-diagram 
estimate at the one loop level; it is easy to realize that
they are proportional to $1/M^2$, which, in units of $\Lambda$,
is a big number. Thus the large values in 
Fig.~\ref{fig:fdue}(a) are just due to the scale employed and, since 
these four couplings for any practical purpose, must 
be compared to the energy scale at which they are calculated, it is 
physically significant to plot them in units of 
the running cutoff $k$:
the corresponding curves are displayed in Fig.~\ref{fig:fdue}(b); 
in this case the couplings are strongly suppressed below $M$.
\par
\begin{figure}
\psfig{figure=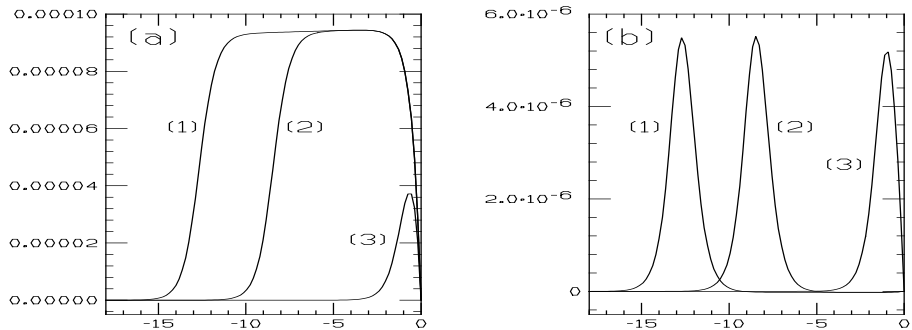,height=4.3cm,width=12.cm,angle=90}
\caption{
(a): $G_{6,0}(t)\cdot k^2$ vs $t$ 
for $M/\Lambda \sim 2\cdot 10^{-6}$ (1), 
$\sim  10^{-4}$ (2), 
$\sim 0.33$ (3).\break
(b): $H_{2,0}(t)\cdot k^2$ for the three values of $M/\Lambda$
quoted in (a). 
\label{fig:ftre}}
\end{figure}

 
It must be remarked that there is no change in the couplings when 
the light mass threshold is crossed. This is a consequence of having 
integrated the heavy field only: in this case one could check directly
from the equations ruling the coupling constants flow, that 
a shift in the 
initial value $G_{2,0}(t=0)$ has the only effect 
(as long as one remains in the symmetric phase)
of modifying $G_{2,0}(t)$, leaving the other curves unchanged.
Therefore the results obtained are independent of $m$ and
do not change even if $m$ becomes much larger than $M$.
An example of the heavy mass dependence is shown in Fig.~\ref{fig:ftre}(a), 
where $G_{6,0}(t)$ is plotted, in units of the running cutoff $k$, for three 
different values of $G_{0,2}(t=0)$, which correspond respectively to 
$M/\Lambda\sim 2\cdot 10^{-6}$, (1), 
$M/\Lambda\sim 10^{-4}$, (2) and  
$M/\Lambda\sim 0.33$, (3). 
Note, in each curve, the change of slope when the $M$ scale is crossed.
$H_{0,0}=1,~~H_{0,2}=0$ is a constant solution of the corresponding equations
for these two couplings; on the other hand $H_{2,0}$ is not constant 
and it is plotted in units of the running cutoff $k$ in Fig.~\ref{fig:ftre}(b),
for the three values of $M$ quoted above.
\par 
In conclusion, according to the AC theorem all couplings are constant at low 
energies 
%(well below $M$) 
and a change in the UV physics can only shift their values in the IR region.
Remarkably, for increasing $t$, 
no trace of UV physics shows up until one reaches $M$, 
that acts as a UV cut-off for the low energy physics.
Moreover, below $M$, no non-perturbative effect appears due to the 
non-renormalizable couplings that vanish fastly in units of $k$.
Their behavior  above $M$ is somehow constrained by the 
renormalizability condition fixed at $t=0$, as clearly shown in
Fig.~\ref{fig:ftre}(a) (3). Finally, the peak of $H_{2,0}$ at $k\sim M$
in Fig.~\ref{fig:ftre}(b), whose width and
height are practically unchanged in the three examples,
is a signal of non-locality of the theory limited to the region 
around $M$.
\section*{Acknowledgments}
A.B. gratefully acknowledges Fondazione A. Della Riccia and INFN 
for financial support.


\section*{References}
\begin{thebibliography}{99}
\bibitem{ac} T.D. Appelquist and J. Carazzone, \Journal{\PRD}{11}{2856}{1975}.
\bibitem{girar}L. Girardello and A. Zaffaroni, \Journal{\NPB}{424}{219}{1994}.
\bibitem{weg}F.J. Wegner and A. Houghton, 
\Journal{{\em Phys. Rev.} A}{8}{401}{1972}.
\bibitem{jan}S.B. Liao and J. Polonyi, \Journal{\PRD}{51}{4474}{1995}.
\end{thebibliography}
\end{document}